# Gate-tunable spin Hall effect in an all-light-element heterostructure: graphene with copper oxide

*Haozhe Yang[1], Maider Ormaza[2], Zhendong Chi[1], Eoin Dolan[1], Josep Ingla-Aynés[1], C.K. Safeer[1], Franz Herling[1], Nerea Ontoso[1], Marco Gobbi[1,3,4], Beatriz Martín-García[1,3], Frederik Schiller[4,5], Luis E. Hueso[1,3], Fèlix Casanova[1,3]\**

[1] CIC nanoGUNE BRTA, 20018 Donostia-San Sebastian, Basque Country, Spain.
[2] Departamento de Polímeros y Materiales Avanzados: Física Química y Tecnología Facultad de Químicas, UPV/EHU, 20080 Donostia-San Sebastián, Basque Country, Spain
[3] IKERBASQUE, Basque Foundation for Science, 48009 Bilbao, Basque Country, Spain.
[4] Centro de Física de Materiales (CSIC-EHU/UPV) and Materials Physics Center (MPC), 20018 Donostia-San Sebastian, Basque Country, Spain.
[5] Donostia International Physics Center, 20018 Donostia-San Sebastian, Basque Country, Spain

ABSTRACT

Graphene is a light material for long-distance spin transport due to its low spin-orbit coupling, which at the same time is the main drawback to exhibit a sizeable spin Hall effect. Decoration by light atoms has been predicted to enhance the spin Hall angle in graphene while retaining a long spin diffusion length. Here, we combine a light metal oxide (oxidized Cu) with graphene to induce the spin Hall effect. Its efficiency, given by the product of the spin Hall angle and the spin diffusion length, can be tuned with the Fermi level position, exhibiting a maximum (1.8 ± 0.6 nm at 100 K) around the charge neutrality point. This all-light-element heterostructure shows a larger efficiency than conventional spin Hall materials. The gate-tunable spin Hall effect is observed up to room temperature. Our experimental demonstration provides an efficient spin-to-charge conversion system free from heavy metals and compatible with large-scale fabrication.

Generation, manipulation, and transport of spin currents is a longstanding topic of interest in spintronics[1–3]. A very convenient way to create (and detect) such spin currents is by exploiting spin-charge interconversion phenomena, such as the spin Hall effect (SHE)[4] or the Edelstein effect (EE)[5,6] and their reciprocal effects. They usually require spin-orbit coupling (SOC)[4–6], a relativistic effect which arises from the effective magnetic field felt by the electron in its rest frame, due to the electrostatic potential of the positive nucleus. SOC increases with the atomic number Z (with a $Z^4$ dependence based on the hydrogen atom model approximation[7]). Most materials investigated for spin-charge interconversion contain high-Z elements such as heavy metals[4] or topological insulators[8], although low-Z elements have recently drawn attention for the potential use of their orbital angular momentum[9,10]. Additionally, the recent focus on the large spin momentum-locking in heterostructures and interfaces[11,12] indicates the critical role of the interfacial effect in condensed matter systems[7,13–15].

Van der Waals materials, which can be exfoliated down to the single atomic thickness and can be combined by stacking them into heterostructures[16], provide a platform for investigating spin-charge interconversion from both basic and applied perspectives[17,18]. As the prototypical van der Waals material, graphene exhibits high mobility, gate tunability, and a long spin lifetime[19], being thus considered one of the best materials for spin transport[20]. However, the low intrinsic SOC makes pristine graphene unsuitable for spin-charge interconversion. One solution is to induce SOC in graphene by proximity with a transition-metal dichalcogenide (TMD), a system which was predicted to show SHE[21–24] and EE[23–25], and was later confirmed experimentally[26–29]. TMD-proximitized graphene shows a sizeable spin Hall angle ($\theta_{SH}$) while preserving a long spin diffusion length ($\lambda_s$)[26–28]. This unique combination provides a large spin-to-charge conversion length ($\theta_{SH}\lambda_s$), which is considered an essential figure of merit for spin-orbit-based devices[30], such as the magnetoelectric spin-orbit (MESO) logic[31]. However, scaling up TMD/graphene van der Waals heterostructures for integrated devices is still challenging[32–34], as the deterministic mechanical transfer technique required to keep a clean interface[35,36] limits the number of devices per chip.

An alternative route to enhance SOC in graphene is to decorate it with adatoms[37–41], which is possible to integrate into a large-scale fabrication process. High-Z atoms can induce SOC in graphene[37–39], and have been observed to exhibit SHE[42]. Low-Z atoms, however, are also predicted to induce SOC in graphene[40,41]. One representative element is copper (Cu): a low-cost light metal which has also been widely used as the catalytic substrate for large-scale CVD graphene growth[43,44]. Cu shows a weaker SOC compared with high-Z elements, although recent studies unveiled that charge-to-spin conversion in Cu can be enhanced with ambient oxidation[45–47]. Furthermore, theoretical works have demonstrated that Cu can induce SOC both intrinsically by proximity effect[41] and extrinsically by adatom decoration[40]. This extrinsic mechanism has been predicted to induce SHE by skew scattering[48,49], with tunable $\theta_{SH}$ by changing the Fermi level position due to the preservation of the 2D Dirac cone (linear dispersion in the band structure). Even though the observation of SHE in graphene decorated with metallic adatoms has been reported using a non-local Hall bar[50], such configuration is prone to spurious effects[49,51–54] and an unambiguous demonstration using spin precession with polarization-selective electrodes is lacking.

In this letter, we report the first experimental observation of the SHE and its inverse in a graphene/oxidized copper ($CuO_x$) heterostructure by measuring spin precession in a lateral spin valve. The Cu layer is grown on the graphene layer without damaging its crystal structure. Moreover, the charge-to-spin conversion output voltage can be tuned by electrical gating. By analyzing our results, we confirm that $\theta_{SH}$ and $\theta_{SH}\lambda_s$ are also gate tunable, showing a maximum value around the charge neutrality point (CNP) of graphene, in agreement with the theoretical predictions[48,49]. Moreover, the gate-tunable SHE can be observed up to room temperature, making this system an ideal candidate for large scale fabrication.

The SHE present at a $CuO_x$/graphene heterostructure is illustrated in Figure 1(a). Because of the 2D nature of graphene, by applying an in-plane current ($I_c$) along $y$, the SHE generates a transverse in-plane spin current ($I_s$) along $x$ with out-of-plane spin polarization. $I_s$ propagates through the adjacent pristine graphene arm and can be non-locally detected by a ferromagnetic (FM) electrode. The reciprocal effect, inverse spin Hall effect (ISHE), converts $I_s$ into a transverse $I_c$. We thus designed and fabricated a device that can detect the spin-to-charge and charge-to-spin conversion by combining a graphene Hall bar (with $CuO_x$ on top of the cross-junction, and arms of pristine graphene) with a FM electrode (see Figure 1b). Adjacent pairs of FM electrodes are used to calibrate the spin injection efficiency of the FM contacts and the spin transport properties of the pristine graphene, required to quantify $\theta_{SH}$. Details of the device fabrication are given in Supporting Information Note 1.

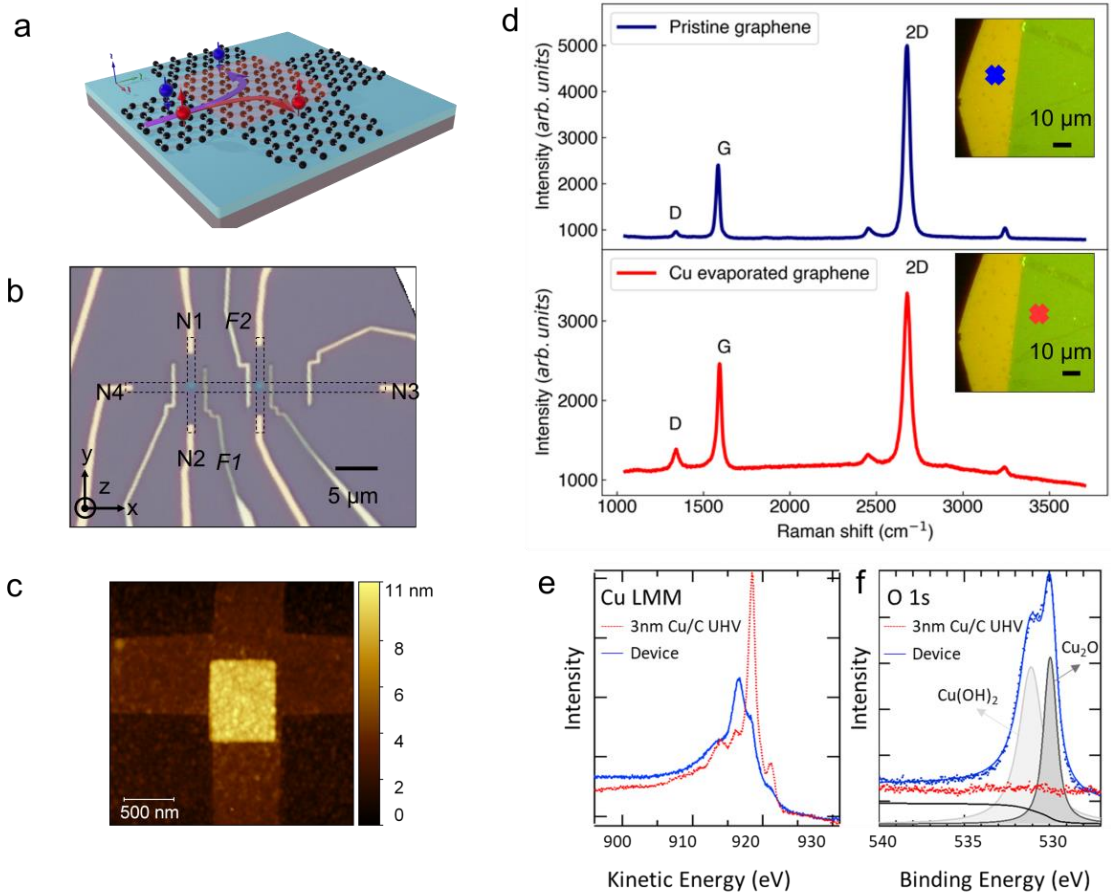

**Figure 1.** (a) Sketch of the spin Hall effect expected in $CuO_x$-covered graphene: the spin current is generated transverse to the charge current direction, with out-of-plane spin polarization. (b) Optical image of Sample 1. The dashed lines follow the edges of the graphene channel. The non-magnetic contact (N1 to N4) and magnetic electrodes (F1 and F2) are indicated. (c) AFM image of the heterostructure at the graphene Hall cross-junction shows a continuous topography of the $CuO_x$ layer. (d) Raman spectra for pristine graphene (upper panel) and the same graphene covered with evaporated $CuO_x$ (lower panel) indicate no clear damage induced by the evaporation. The insets show the optical image of the pristine graphene region (yellow) and the region covered by $CuO_x$ (green), with an indication of the Raman laser spot position. (e) Cu LMM Auger spectra and (f) O 1s spectra for a 3-nm-thick Cu film grown on top of graphite under UHV conditions (in red) and for a 5-nm-thick Cu film grown on graphite and left in air for 72 h prior to measurement, following the same protocol as the device fabrication (in blue). Fitting of the latter shows two main contributions, one from $Cu_2O$ and another one from $Cu(OH)_2$.

Cu has low affinity to carbon (C) and does not form any carbide phases[55,56]. It is considered to form only soft bonds with C via charge transfer from the π electrons in the sp$^2$ hybridized C to the empty 4s states of Cu[44]. Before measuring the devices, we performed several characterizations. We first checked whether graphene is degraded after thermally evaporating Cu on top and leaving it to oxidize. The control experiment was performed on a CVD-grown single layer graphene, in which a selected area is covered by 5 nm of thermally evaporated Cu, followed by ambient oxidization. The Raman spectra of the evaporated/non-evaporated areas are plotted in Figure 1(d), both showing well-defined 2D and G peaks, indicating no structural damage[57] of graphene when forming the heterostructure. To characterize the electrical properties of $CuO_x$, we fabricated a double Hall bar with 5-nm-thick Cu grown and oxidized in the same conditions and measured its charge transport properties, indicating that the conductivity of $CuO_x$ is at least seven orders of magnitude smaller than that of graphene (see Supporting Information Figure S7). This confirms that all charge transport (and thus spin transport) in our devices only occurs in graphene. The atomic force microscopy (AFM) image in Figure 1(c) shows the topography of the heterostructure, exhibiting a continuous $CuO_x$ film with root mean square roughness of 0.6 nm, much smaller than the thickness.

X-ray photoelectron spectroscopy (XPS) was employed to unveil the chemical state of Cu on the graphene surface. Figure 1(e) shows the comparison between the Cu-LMM Auger transition of a 3-nm-thick Cu grown on graphite in ultra-high vacuum (UHV) and a 5-nm-thick Cu film grown on graphite after 72 h of air exposure. The former shows metallic state, while the latter, which is analogous to the one employed to fabricate the devices, shows a complete oxidation[58]. Furthermore, the O 1s spectrum of the 5-nm-thick film in Figure 1(f) shows two main contributions, one related to $Cu_2O$ and another to $Cu(OH)_2$, with binding energies 530.0 eV and 531.1 eV, respectively, whereas the O 1s peak vanishes for the 3-nm-thick Cu grown in UHV. No traces of C-O or Cu-C bonds have been found on the C 1s spectrum, as shown in Supporting Information Note 3. Additional XPS measurements have been carefully performed to further investigate the oxidation process of the Cu on the surface. From this analysis, we conclude that the amount of metallic Cu on the graphene surface is negligible.

After confirming that graphene is not structurally damaged, Cu is fully oxidized and that $CuO_x$ and $Cu(OH)_2$ are not conductive and fully cover graphene, we proceed to study the spin transport properties of our devices. First, we use two FM electrodes as the spin injector (F1) and detector (F2) to extract the spin transport properties of the pristine graphene (see Figure 2(a)). A charge current $I_c$ is applied from F1 to N4, and a non-local voltage $V_{NL}$ is measured between F2 and N3. We normalized $V_{NL}$ by $I_c$ to obtain a non-local resistance $R_{NL}$. The relative alignment of the magnetization of F1 and F2 electrodes (parallel or antiparallel) can be set with an external magnetic field ($B_y$) applied along their easy axis $y$, because the different widths of F1 and F2 yield different coercivities. A magnetic field ($B_x$) along the in-plane hard axis $x$ of the FM contacts is swept from zero until full saturation of the FM electrodes. At lower fields, spins in graphene precess in the $y$-$z$ plane and $R_{NL}$ exhibits a symmetric Hanle precession behavior. One representative result measured at 100 K with a back gate voltage ($V_g$) of 30 V is shown in Figure 2(b). The pure spin precession signal $\Delta R_{NL}$ plotted in Figure 2(c) is obtained by subtracting the $R_{NL}$ curves between the parallel ($R_{NL}^P$) and antiparallel ($R_{NL}^{AP}$) configuration.

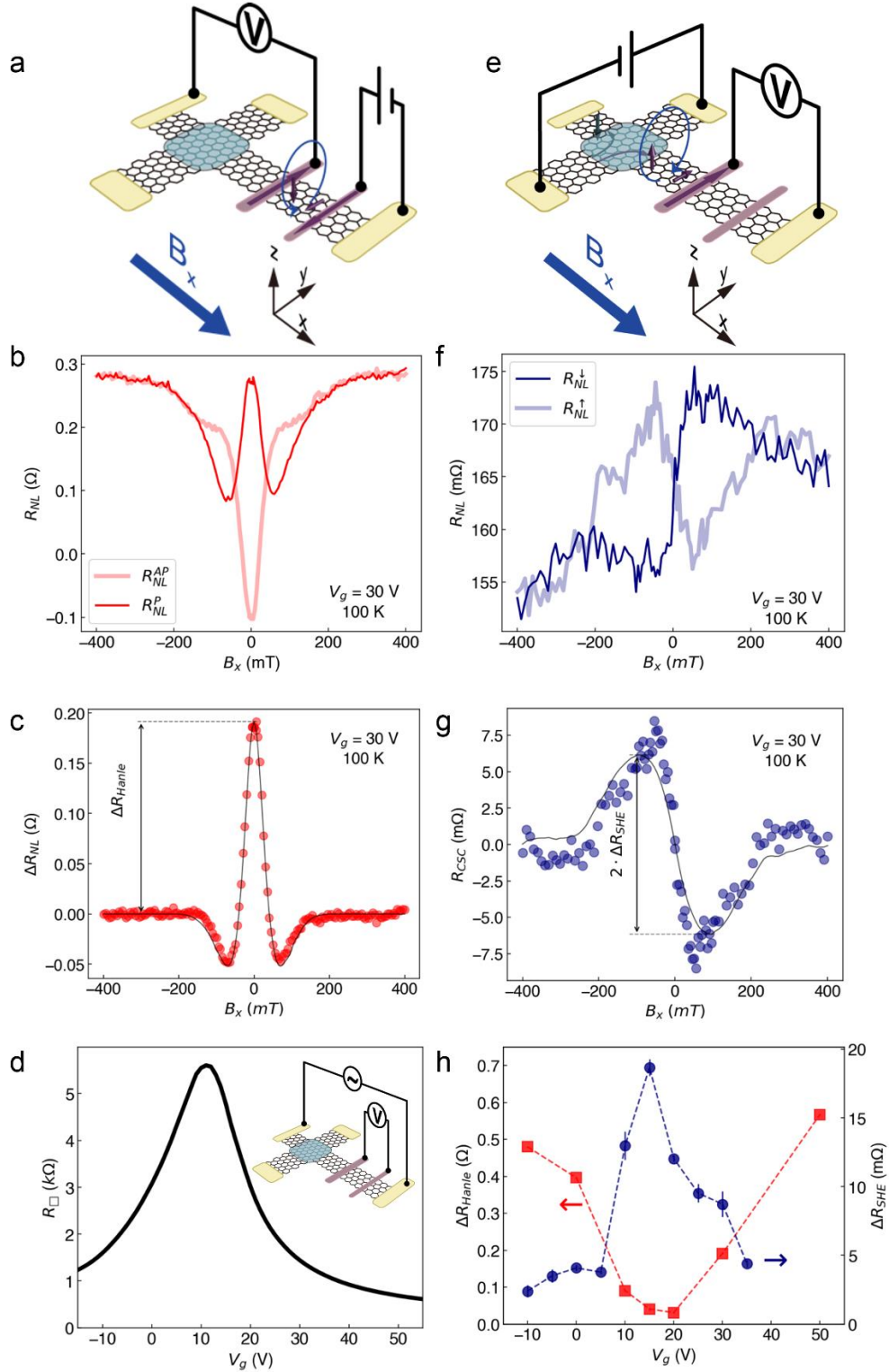

**Figure 2.** (a) Measurement configuration for symmetric Hanle precession with standard electrical spin injection and detection using two FM electrodes. An in-plane magnetic field, $B_x$, is applied to induce precession of the $y$-polarized spins injected from the FM electrode into the graphene. (b) Non-local resistance as a function of $B_x$ measured at 100

K and $V_g$ = 30 V using the configuration in (a), with F1 and F2 electrodes set in a parallel ($R_{NL}^P$, dark-red line) and antiparallel ($R_{NL}^P$, light-red line) configuration. (c) Net symmetric Hanle precession signal extracted from the two curves in (b) by taking $\Delta R_{NL} = (R_{NL}^P - R_{NL}^{AP})/2$. The gray solid line is a fit of the data to the solution of the Bloch equation. The amplitude of the signal, $\Delta R_{Hanle}$, is labeled. (d) Square resistance, $R_\square$, of the pristine graphene as a function of $V_g$ measured at 100 K using the four-point configuration illustrated in the inset. (e) Measurement configuration for antisymmetric Hanle precession with charge-to-spin conversion at the $CuO_x$-covered graphene and detection using a FM electrode. $B_x$ is applied to induce precession of the z-polarized spins, originating from the SHE in decorated graphene. (f) Non-local resistance as a function of $B_x$ measured at 100 K and $V_g$ = 30 V using the configuration in (e), with initial positive ($R_{NL}^\uparrow$, light-blue line) and negative ($R_{NL}^\downarrow$, dark-blue line) magnetization direction of F1. (g) Net antisymmetric Hanle precession signal extracted from the two curves in (f) by taking $R_{CSC} = (R_{NL}^\uparrow - R_{NL}^\downarrow)/2$ and antisymmetrizing it. The gray solid line is a fit of the data to the solution of the Bloch equation. The amplitude of the signal, $\Delta R_{SHE}$, is labeled. (h) Amplitude of the symmetric ($\Delta R_{Hanle}$, blue squares) and antisymmetric ($\Delta R_{SHE}$, red circles) Hanle precession signals at 100 K plotted as a function of $V_g$. All data corresponds to Sample 1.

After confirming the spin transport in our device, we perform the charge-to-spin conversion measurements (see Figure 2(e)). $I_c$ is applied along the transverse channel *y* from N1 to N2. At the $CuO_x$/graphene heterostructure, the SHE generates an $I_s$ along *x* with an out-of-plane spin polarization, which diffuses into the pristine graphene channel. By applying $B_x$, the out-of-plane spins precess in the *y-z* plane. Due to this precession, the diffusing spins develop a *y* component which can then be detected by F1, which is magnetized along *y*, as a non-local voltage (again normalized into a non-local resistance, $R_{NL}^\uparrow = V_{NL}/I_c$). This process results in an antisymmetric spin precession curve with a maximum and a minimum at a certain $\pm B_x$ value. When reversing the magnetization of F1, the antisymmetric Hanle curve $R_{NL}^\downarrow$ is reversed because the detector senses the opposite *y*-spin component. To obtain the two curves, we initialize the F1 magnetization with a $B_y$ field, then sweep $B_x$ from zero until saturation of the electrode. This operation is repeated for the two polarizations and $B_x$ polarities. One representative charge-to-spin conversion curve set measured at 100 K and $V_g$ = 30 V is shown in Figure 2(f), exhibiting the expected behavior. The pure charge-to-spin precession signal $\Delta R_{CSC}$ with out-of-plane spin polarization (due to the SHE), plotted in Figure 2(g), has been obtained as follows: (1) subtracting the $R_{NL}$ curves between the positive ($R_{NL}^\uparrow$) and negative ($R_{NL}^\downarrow$) alignment of the F1 magnetization, followed by (2) an antisymmetrization of the obtained curve with respect to the magnetic field. The first step removes any initial-magnetization independent components, such as local magnetoresistance, ordinary Hall effect[59] and conventional EE[26,60]. The second step eliminates the possible contribution of unconventional charge-to-spin conversion with spins polarized along *y*, which has been observed in some van der Waals heterostructures[60–63]. The reciprocal experiment (inverse SHE) is shown in Supporting Information Figure S11 and confirms that we are in the linear response regime. A control experiment in a reference device without $CuO_x$ as an adlayer exhibits no spin-to-charge signal (see Supporting Information Figure S10).

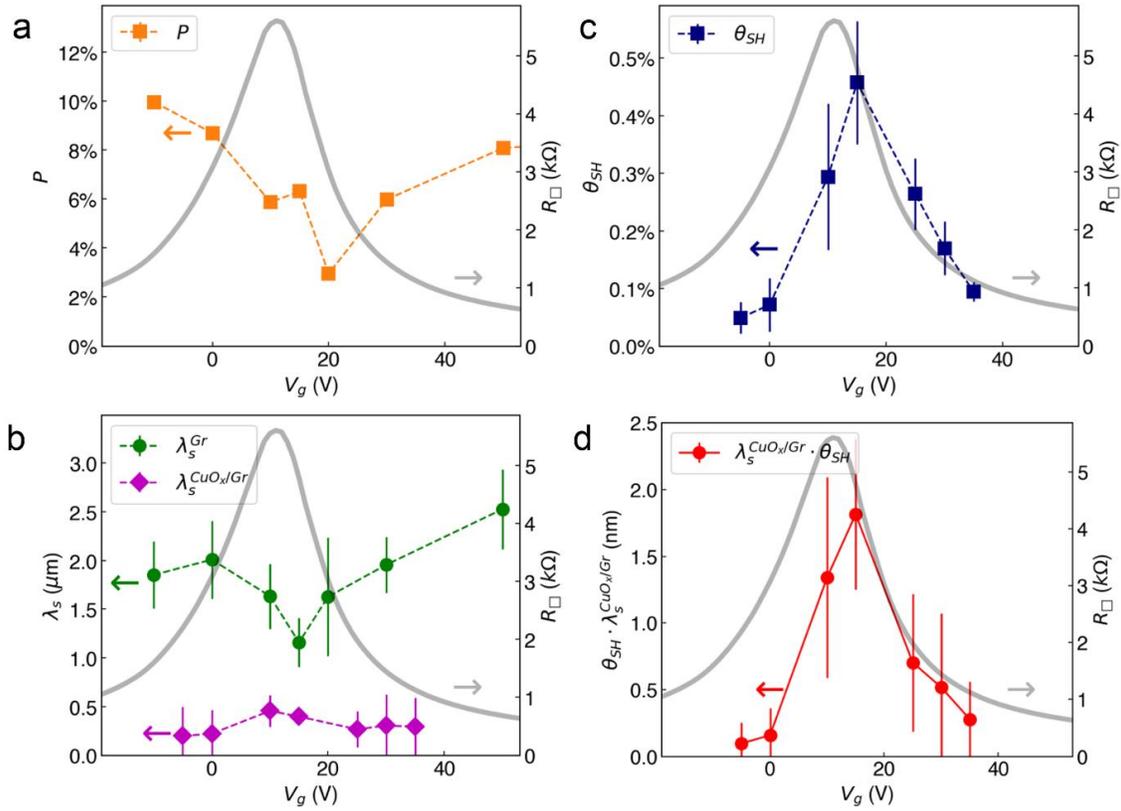

**Figure 3.** Gate voltage dependence of (a) spin polarization of Co/TiO$_x$ electrode, (b) spin diffusion length of pristine graphene (green circles) and CuO$_x$/graphene (magenta diamonds), (c) spin Hall angle of CuO$_x$/graphene, and (d) spin-to-charge conversion length of CuO$_x$/graphene. All results are taken from Sample 1 at 100 K. The square resistance is plotted in all panels as a solid gray line for comparison.

Next, we studied how the charge-to-spin conversion is tuned with $V_g$. We first measured the charge transport properties with a four-point configuration. The gate-dependent square resistance ($R_\square$) of pristine graphene measured at 100 K is plotted in Figure 2(d), showing a representative Dirac material feature, with the resistance maximum corresponding to the CNP. The CuO$_x$-covered graphene and the pristine graphene exhibit a similar doping, as shown in Supporting Information Figure S12. ARPES experiments have shown that graphene preserves its Dirac cone in contact with copper oxide[64], which is promising for gate-tunable electronics. Then, we measured the symmetric and antisymmetric Hanle precession curves at different $V_g$, taking the gate-dependent spin injection efficiency and spin transport properties of the pristine graphene[65]. The amplitude $\Delta R_{Hanle}$ of the symmetric Hanle curve (defined in Figure 2(c)) and the amplitude $\Delta R_{SHE}$ of the antisymmetric Hanle curve (defined in Figure 2(g)) are plotted in Figure 2(h). Here $\Delta R_{Hanle}$ and $\Delta R_{SHE}$ exhibit the opposite behavior with $V_g$. $\Delta R_{Hanle}$ shows a minimum around the CNP, increasing when the carrier density is higher. Such decrease of the spin signal around the CNP is expected when the FM/graphene contacts have a low contact resistance ($R_c$) due to the spin conductivity mismatch[66], which is our case ($R_c$=1.7 kΩ and 3 kΩ estimated from a three-point measurement). In stark contrast, $\Delta R_{SHE}$ shows the highest value around the CNP, where the spin injection efficiency is lowest, implying that the charge-to-spin conversion efficiency should be larger around the CNP to achieve an overall larger output signal.

For a quantitative analysis of the spin transport and the charge-to-spin conversion efficiency, both the symmetric and antisymmetric Hanle precession data are fitted to the numerical solution of the Bloch equation (see solid gray lines in Figs. 2(c) and 2(g)). The fitting procedure is detailed in Supporting Information Note 2. From the fit of the symmetric Hanle curve, we obtain the spin polarization $P$ of the FM contact and the spin lifetime, $\tau_s^{gr}$, of the pristine graphene. In order to decrease the number of fitting parameters, we assume the spin and charge diffusion constants to be equal ($D_s = D_c$)[66] and extract $D_c$ from $R_\square$. The gate dependence of $P$ is plotted in Figure 3(a), with values ranging from 9.9 ± 0.1% at $V_g = -10$ V down to 3.0 ± 0.2% at $V_g = 20$ V, close to the CNP. The spin diffusion length $\lambda_s^{gr} = \sqrt{\tau_s^{gr} D_s}$ of the pristine graphene is plotted in Figure 3(b), with values ranging from 2.5 ± 0.4 µm at $V_g = 50$ V to a minimum value of 1.2 ± 0.2 µm at $V_g = 15$ V, close to the CNP.

From the fit of the antisymmetric Hanle curve, and by fixing $P$ and $\tau_s^{gr}$ obtained from the previous fit, we obtain the spin Hall angle $\theta_{SH}$ and the spin lifetime $\tau_s^{CuOx/gr}$ of the decorated graphene. Note that no evidential spin lifetime anisotropy of the functionalized graphene was observed, shown in Supporting Information Figure S14. The gate dependence of $\theta_{SH}$ is plotted in Figure 3(c), showing a peak of 0.5 ± 0.1% around the CNP and decreasing away from the CNP. Resonant skew scattering has been theoretically predicted in atomic-decorated graphene, exhibiting a spin Hall angle which is tuned with the Fermi level position and the largest value expected near the CNP[48,49], in excellent agreement with our observation. This extrinsic mechanism is therefore the most likely scenario in our system.

Note that the obtained value of $\theta_{SH}$ is even larger than that of BiO$_x$/graphene at the same temperature (~0.25% at 100 K), although Bi has a much larger $Z$, and thus SOC, than Cu. The low affinity and ability to form soft bonds with carbon[44], discussed previously, makes Cu not only an excellent catalyst for graphitic carbon formation but could also induce SOC by proximity. Considering the well-preserved crystal structure of graphene shown by Raman spectroscopy in Figure 1(d) and the continuous topography of the CuO$_x$ layer grown on graphene shown in the AFM image in Figure 1(c), a long-range flat interface is possible to be formed, which is an ideal proximitized system. In this case, we cannot rule out that a spin-orbit proximity effect such as the one calculated in Ref. [41] is able to provide an extra contribution to our large $\theta_{SH}$ in the CuO$_x$-covered graphene.

The spin diffusion length of the CuO$_x$/graphene heterostructure is calculated as $\lambda_s^{CuOx/gr} = \sqrt{\tau_s^{CuOx/gr} D_s}$ and is plotted in Figure 3(c). $\lambda_s^{CuOx/gr}$ is substantially shorter (ranging between 450 ± 160 nm at $V_g = 10$ V and 190 ± 300 nm at $V_g = -5$ V) than $\lambda_s^{gr}$, which is expected because an enhanced SOC decreases the spin lifetime. The gate-dependent spin-to-charge conversion length ($\theta_{SH} \lambda_s^{CuOx/gr}$), plotted in Figure 3(d), also exhibits a peak around the CNP. At $V_g = 15$ V, $\theta_{SH} \lambda_s^{CuOx/gr} = 1.80 ± 0.56$ nm, which is one order of magnitude larger than the value at $V_g = 0$ V (0.15 ± 0.20 nm). This is a remarkable result: while we have only low-Z elements in our material system (Cu, O, C), the value is higher than in conventional large SOC materials, such as heavy metals (0.2 nm for Pt[67], 0.34 nm for W[68]) or metallic Rashba interfaces (0.3 nm for Ag/Bi[69], −0.17 nm for Cu/Au[70]), and is of the same order of magnitude as in material systems such as topological insulators (2.1 nm for α-Sn[71]), oxide 2DEGs (6.4 nm for LAO/STO[11]), or heavy metal oxide/graphene heterostructures (1 nm for $BiO_x$/graphene at 100 K[42]).

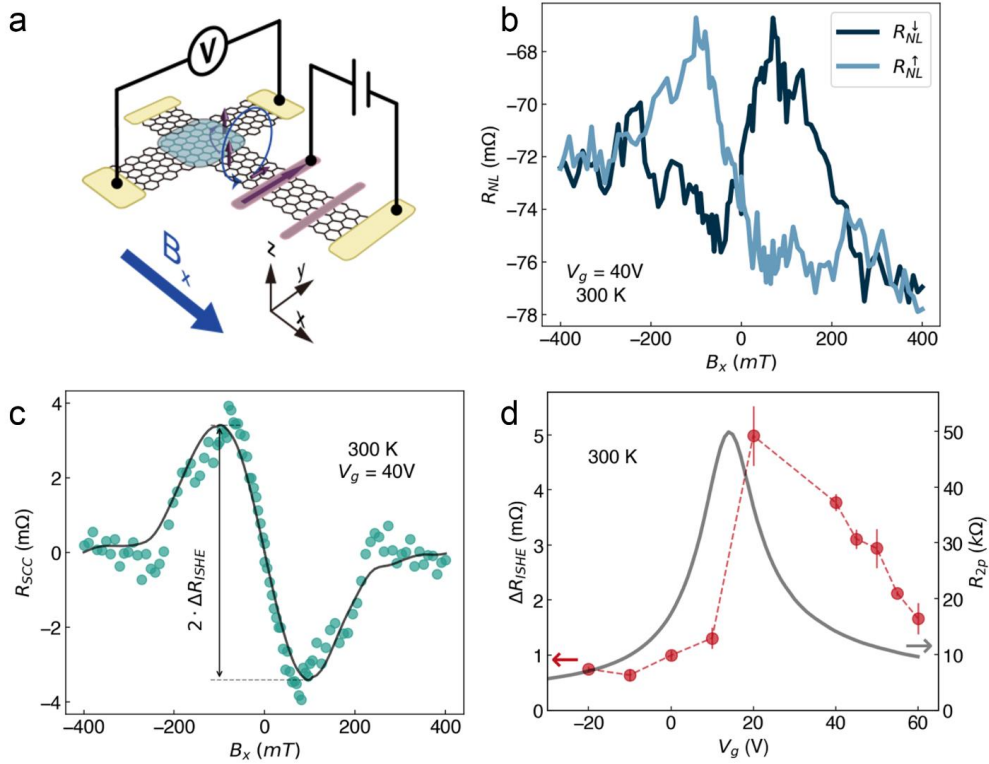

**Figure 4.** (a) Measurement configuration for antisymmetric Hanle precession with spin injection using a FM electrode and spin-to-charge conversion at the $CuO_x$-covered graphene. $B_x$ is applied to induce precession of the $y$-polarized spins towards $z$, in order to be detected by ISHE in decorated graphene. (b) Non-local resistance as a function of $B_x$ measured at 300 K and $V_g$ = 40 V in Sample 2 using the configuration in (a) with initial positive ($R_{NL}^\uparrow$, light-blue line) and negative ($R_{NL}^\downarrow$, dark-blue line) magnetization direction of the FM injector. (c) Net antisymmetric precession signal extracted from the two curves in (b) by taking $R_{SSC} = (R_{NL}^\uparrow - R_{NL}^\downarrow)/2$ and antisymmetrizing it. The gray solid line is a fit of the data to the solution of the Bloch equation. The amplitude of the signal, $\Delta R_{ISHE}$, is labeled. (d) Gate dependence of $\Delta R_{ISHE}$ at 300 K (solid red circles). The two-point resistance as a function of $V_g$ of the $CuO_x$-covered graphene at the cross-junction of the Hall bar measured at 300 K is also plotted (solid gray line).

Finally, we studied the spin-to-charge conversion of $CuO_x$/graphene heterostructure at room temperature (see Figure 4(a)). The non-local spin precession measurement measured at 300 K and $V_g$ = 40 V in Sample 2 is exemplary shown in Figure 4(b). Sample 2 is fabricated on the same single layer graphene flake as Sample 1, with the details shown in Supporting Information Figure S9. The pure spin-to-charge precession signal $R_{SCC}$ due to the ISHE, obtained in the same way as for $R_{CSC}$, is plotted in Figure 4(c) and shows a clear antisymmetric Hanle behavior. We also observed that its amplitude $\Delta R_{ISHE}$ could be tuned with $V_g$, exhibiting the largest output at 20 V, as shown by solid red circles in Figure 4(d). The solid gray line in Figure 4(d) is the two-point resistance measurement of the transverse Hall arm at room temperature, indicating the CNP to be around 15 V, close to the $\Delta R_{ISHE}$ maximum. An estimation of $\theta_{SH}$ and $\lambda_s^{CuOx/gr}$ at room temperature requires knowing $P$ and $\lambda_s^{gr}$ at this temperature. However, the shorter $\lambda_s^{gr}$ at 300 K decreases the signal for the symmetric Hanle precession, yielding a low signal-to-noise ratio that leads to an unreliable estimation of the spin transport parameters. We therefore use $P$ and $\lambda_s^{gr}$ values obtained at 100 K for Sample 1 to estimate $\theta_{SH}$ and $\lambda_s^{CuOx/gr}$ at room temperature. With this approximation, we would obtain $\theta_{SH}$ = 0.34 ± 0.18% and $\lambda_s^{CuOx/gr}$ = 370 ± 430 nm at room temperature and $V_g$ = 20 V, corresponding to a $\theta_{SH}\lambda_s^{CuOx/gr}$ of 1.2 ± 1.6 nm. The robustness of this effect is further confirmed by additional samples (including trilayer graphene) shown in Supporting Information Figure S13.

In conclusion, we report the first unambiguous experimental observation of the SHE in graphene induced by a light-metal oxide. The $CuO_x$-covered graphene keeps its structural and electrical properties, while acquiring SOC from the oxidized Cu adlayer that leads to SHE up to room temperature. After a careful estimation considering the gate-dependent spin transport properties of the pristine graphene and the FM contacts, we find that the spin Hall angle and the spin-to-charge conversion length of $CuO_x$/graphene are gate tunable, with maximum values around the CNP. These values are comparable to those observed in heavy metal oxide/graphene heterostructures. Cu being a well-established light element for the graphene industry, $CuO_x$/graphene heterostructures could be an ideal candidate for large-scale spintronic applications.

ASSOCIATED CONTENT
**Supporting Information**.
Additional details on methods, spin precession fitting process, XPS analysis, electrical characterization of oxidized copper, optical image of devices during the fabrication process, optical and SEM images of Sample 2, reference non-local Hall bar without CuOx, reciprocity between spin Hall effect and inverse spin Hall effect, charge transport properties of the pristine graphene and the CuOx-covered graphene, reproducibility.

AUTHOR INFORMATION
**Corresponding Author**
*E-mail: f.casanova@nanogune.eu
**Author Contributions**
H.Y. and F.C. conceived the study. H.Y. performed the sample fabrication, electrical, and optical measurements, with the help of Z.C., E.D., J.I.A, C.K.S., F.H., N.O. and B. M.-G.. M.O. and F.S. performed the XPS experiments. All authors contributed to the discussion of the results and their interpretation. H.Y. and F.C. wrote the manuscript with input from all authors.

ACKNOWLEDGMENTS
We acknowledge funding by the "Valleytronics" Intel Science Technology Center, by the Spanish MICINN (Project No. PID2021-122511OB-I00 and "Maria de Maeztu" Units of Excellence Programme No. CEX2020-001038-M), by the European Union H2020 under the Marie Sklodowska–Curie Actions (Project Nos. 0766025-QuESTech and 955671-SPEAR) and by Diputación de Gipuzkoa (Project No. 2021-CIEN-000037-01). Z.C. acknowledges postdoctoral fellowship support from "Juan de la Cierva" Programme by the Spanish MICINN (grant No. FJC2021-047257-I).

# Supporting Information

**Note 1 Methods.**

**Sample fabrication.** Single layer graphene was exfoliated from bulk graphite crystals (supplied by NGS Naturgraphit GmbH) on Si substrates with 300 nm $SiO_2$. Calibrated optical microscopy was used to identify the number of layers. The device was then fabricated with 4 e-beam lithography steps. First, the graphene was patterned into several Hall bars with a width of 500 nm using e-beam lithography. 20-nm-thick Al deposited by high-vacuum thermal evaporation (base pressure $\sim 7\times10^{-7}$ torr) was used as a hard mask, then the graphene was reactive-ion etched using an $Ar/O_2$ plasma. The Al hard mask was removed with a base developer solution with tetra-methyl ammonium hydroxide, followed by annealing of the device at 400°C for 1 hour in ultra-high vacuum ($\sim 2\times10^{-8}$ torr) to remove residues from the graphene. Second, a 5-nm-thick Cu layer was deposited in the cross-junction of the graphene Hall bar using e-beam lithography, ultra-high vacuum thermal evaporation (base pressure $< 1.2\times10^{-9}$ torr), and lift-off process. The device was left in atmosphere for 72 hours for oxidization. Third, graphene was contacted with non-magnetic Pd(5 nm)/Au(45 nm) contacts fabricated using e-beam lithography followed by e-beam evaporation and lift-off. Finally, the magnetic $TiO_x$/Co electrodes were fabricated by e-beam lithography, deposition of 2.6 Å of Ti (followed by oxidation in air for 10 minutes), 35 nm of Co, and 10 nm of Au as capping layer by e-beam evaporation, and lift-off. The width of the magnetic electrodes are 150 nm or 300 nm in order to have different coercivity. The optical images of the devices during each fabrication step are shown in Fig. S8.

**Optical measurements.** Raman spectroscopy characterization was carried out in an Alpha 300R Confocal Raman WITec microscope using a 532 nm laser (incident power $< 1$ mW to avoid damage to the samples during the Raman spectra acquisition), a diffraction grating of 600 l/mm and 100× objective (N.A. 0.90).

**Electrical measurements.** The transport measurements are performed in a physical property measurement system (PPMS) by Quantum Design using a DC reversal technique with a Keithley 2182 nanovoltmeter and a 6221 current source. The n-doped Si substrate acts as a back-gate electrode to which we apply the gate voltage across 300 nm of $SiO_2$ with a Keithley 2636A.

**Note 2 Spin precession fitting process.**

**1. Symmetric Hanle fitting by considering the low contact resistance of the TiO$_x$/ Co contact.**
To analyze the symmetric Hanle precession experiments obtained while having a parallel and or antiparallel orientation of the Co magnetization, we model the spin propagation in our devices using the Bloch equations:

$$D_s \nabla^2 \vec{\mu} - \frac{\vec{\mu}}{\tau_s} + \vec{\omega} \times \vec{\mu} = 0 \tag{1}$$

where $\vec{\mu}$ is the spin accumulation, $D_s$ the spin diffusion constant, and $\tau_s$ the spin lifetime. $\vec{\omega} = g\mu_B \vec{B}$ is the Larmor frequency, $g=2$ is the Landé factor, $\mu_B$ is the Bohr magneton, and $\vec{B}$ is the applied magnetic field.

The spin precession is induced in the $y - z$ plane when applying a magnetic field along $x$ direction. In this case, Eq. (1) turns into

$$D_s \frac{d^2}{dx^2}\begin{pmatrix}\mu_x \\ \mu_y \\ \mu_z\end{pmatrix} - \begin{pmatrix}\tau_x^{-1} \\ & \tau_y^{-1} \\ & & \tau_z^{-1}\end{pmatrix}\begin{pmatrix}\mu_x \\ \mu_y \\ \mu_z\end{pmatrix} + \omega\begin{pmatrix}0 \\ \mu_y \\ -\mu_z\end{pmatrix} = 0 \tag{2}$$

The solution of $\mu_y$ and $\mu_z$ in Eq. (2) is

$$\mu_y = A e^{\frac{x}{\lambda_s}\sqrt{1+i\omega\tau_s}} + B e^{\frac{x}{\lambda_s}\sqrt{1-i\omega\tau_s}} + C e^{-\frac{x}{\lambda_s}\sqrt{1+i\omega\tau_s}} + D e^{-\frac{x}{\lambda_s}\sqrt{1-i\omega\tau_s}} \tag{3}$$

$$\mu_z = -iA e^{\frac{x}{\lambda_s}\sqrt{1+i\omega\tau_s}} + iB e^{\frac{x}{\lambda_s}\sqrt{1-i\omega\tau_s}} - iC e^{-\frac{x}{\lambda_s}\sqrt{1+i\omega\tau_s}} + iD e^{-\frac{x}{\lambda_s}\sqrt{1-i\omega\tau_s}} \tag{4}$$

where $\lambda_s = \sqrt{\tau_s D_s}$ is the spin diffusion length. $A$, $B$, $C$ and $D$ are the coefficients determined by the boundary conditions. The spin current is defined as:

$$I_{Sy(z)} = -\frac{W_{gr}}{eR_\square}\frac{d\mu_{y(z)}}{dx}, \tag{5}$$

where $W_{gr}$ is the width of the graphene channel and $R_\square$ is the square resistance of graphene. The spin accumulation at the detection position $x = x_{det}$ is then converted into a voltage with

$$V_{det}^S = P\frac{\mu_y(x_{det})}{e}, \tag{6}$$

where $P$ is the spin polarization of the Co detector. We consider that the Co injector has the same spin polarization. $V_{det}^S$ is usually normalized by the charge current $I_c$, giving the non-local resistance

$$R_{NL} = \frac{V_{det}^S}{I_c}. \tag{7}$$

The non-local resistance is also affected by the contact pulling effect[1]. Because of the finite/small in-plane shape anisotropy of the ferromagnetic electrodes, the magnetization is then pulled by an angle β from their easy axis towards the field direction, injecting spins along $x$ and resulting in an additional term to the non-local resistance. The $R_{NL}$ obtained in this case is

$$R_{NL} = \pm R^s cos(\beta_1)cos(\beta_2) + R^{s0} sin(\beta_1)sin(\beta_2). \quad (8)$$

Here ± represents the non-local signal with parallel (P) and antiparallel (AP) configuration of the Co electrode magnetizations. $\beta_1$ and $\beta_2$ correspond to the angle β of the two magnetic electrodes used as injector and detector. For simplicity, we consider that the two Co electrodes have the same pulling effect with the angle, $\beta = \beta_1 = \beta_2$. $R^{s0}$ is a constant that corresponds to the spin signal at zero magnetic field. The net spin signal $\Delta R_{NL}$ is then expressed as:

$$\Delta R_{NL} = \frac{R_{NL}^P - R_{NL}^{AP}}{2} = R^s \cos^2(\beta). \quad (9)$$

Because of the opposite spin precession sign of $R_{NL}^P$ and $R_{NL}^{AP}$, the sum $R_{sum} = R_{NL}^P + R_{NL}^{AP}$ is then proportional to $\sin^2(\beta)$. For our analysis, we used the contact pulling from the measurement points on the curve of $R_{sum}$ vs. $B$.

We thus determine the spin signal using the following boundary conditions:

1. Spin accumulation μ is continuous.
2. Spin current is continuous except:
    a. At F1 with spin injection by the charge current, by $\Delta I_s = I_c \cdot P/2$;
    b. At F1 and F2 due to the spin backflow effect because of the low contact resistance, by $\Delta I_s = -\mu_s/(2eR_c)$, where $R_c$ is the contact resistance.
3. Spin current is zero at the sample ends.

We write these boundary conditions in a matrix $X$ which fulfills: $MX = Y$, where $M$ contains the coefficients $M = A, B, C ...$ and $Y = (0, ..., \frac{I_c P}{2}, 0, ...)$ and use the Moore-Penrose inverse to invert $X^{-1}$ and obtain $M = YX^{-1}$. Using the numeric solution described above, we input the measured net spin signal $\Delta R_{NL}$, square resistance $R_\square$, pulling factor $\cos^2(\beta)$, channel size $W_{gr}$ determined by SEM, and the contact resistance $R_c$.

As our graphene is directly on SiO₂, the small free mean path enables us to neglect the spin diffusivity decrease sourcing from the electron-electron scattering and assume that the spin and charge diffusion coefficients are equal[2–4]. To decrease the number of fitting parameters, therefore, we fix the charge diffusion coefficient ($D_c$), extracted with

$$D_c = \frac{\hbar v_{F0}}{2e^2 R_\square} \sqrt{\frac{\pi}{|n|}}. \quad (10)$$

Here, $v_{F0} = 10^6$ m/s is the Fermi velocity, $\hbar$ is the reduced Planck constant, $n$ is the carrier density determined with

$$n = \epsilon_0 \epsilon_{SiO2}(V_g - V_{CNP})/e \cdot d_{SiO2} \quad (11)$$

with $\epsilon_0$ being the vacuum permittivity, $\epsilon_{SiO2}$ = 3.9 for SiO$_2$ relative dielectric permittivity, $d_{SiO2}$ is the thickness of SiO$_2$ of 300 nm. We then fit the polarization $P$ and the spin lifetime $\tau_s^{Gr}$ of pristine graphene.

## 2. Antisymmetric spin precession fitting for SHE and ISHE signals.

We determine the spin Hall angle $\theta_{SH}$ and the spin lifetime $\tau_s^{CuOx/gr}$ of the graphene in the heterostructure with the antisymmetric spin precession signal $\Delta R_{ISHE}$ originated from the (inverse) spin Hall effect of functionalized graphene. Here, we derive the ISHE but, due to reciprocity, the obtained equation is valid for both configurations. The non-local resistance $\Delta R_{ISHE}$ is given by

$$\Delta R_{ISHE} = \frac{\theta_{SH} R_\square \overline{I_{Sz}}}{I_c} \cdot \cos(\beta) \tag{12}$$

where $\overline{I_{Sz}}$ is the average spin current in the cross-junction area of the Hall bar, which is defined as $\overline{I_{Sz}} = \frac{1}{W_{cr}} \int_L^{L+W_{cr}} I_{Sz}(x) dx$, where $L$ is the distance from F1 electrode to the edge of the Hall cross and $W_{cr}$ is the width of the cross. Cos($\beta$) is the term indicating the contact pulling of the ferromagnetic electrode. Note that, since there is only one magnetic electrode, this term is not squared as in Eq. (9). To determine $I_{Sz}(x)$, we use the equations and boundary conditions described in Section 7.1. Finally, since the CNP is the same for the pristine and CuO$_x$-covered regions (see Fig. S12), we assume the charge diffusion coefficients for the CuO$_x$-covered region and the pristine region to be equal.

**Note 3: XPS analysis.**

X-ray photoemission spectroscopy (XPS) was employed to investigate the oxidation of the 5-nm-thick Cu film used in the device fabrication. In order to judge the behavior observed in the devices, two additional samples were initially prepared under ultra-high vacuum (UHV) conditions, one with a low coverage of 0.8 nm and a second one with a higher coverage of 3 nm of Cu. The XPS of these two samples, which were successively exposed to air, were then compared to the 5-nm-thick Cu film (equivalent to the conditions of the device). The substrate material for Cu deposition was graphite (HOPG), which was exfoliated in the fast-entry chamber ($p = 1 \times 10^{-8}$ mbar) and additionally heated to 400 °C. The low energy electron diffraction (LEED) pattern of the graphite was a circle indicating the existence of multiple graphite domains within the LEED spot area. The XPS reference spectrum showed no trace of oxygen or other impurities (Fig. S2, top panels).

In the next step, pure Cu (99.999% purity) was evaporated at a rate of approximately 0.1 nm per minute. The pressure during evaporation was always kept below $p = 5 \times 10^{-9}$ mbar. After Cu deposition, the samples were transferred without leaving the UHV environment to the XPS chamber.

The evolution of the oxidation of the low coverage sample was investigated by measuring the XPS spectra prior and after different air exposure times: 1 h, 24 h and 72 h. Air exposure means here that the sample was kept in the laboratory at room temperature and at ambient relative humidity of approx. 60%. The high coverage sample was measured before and after 72 h air exposure to directly compare with the 5-nm-thick Cu film prepared at the same conditions as the Cu film employed for device fabrication.

X-ray photoelectron spectroscopy was carried out holding the sample at room temperature and illuminating it with monochromatized Al K$_\alpha$ light (hυ = 1486.6 eV) from a microfocus setup (SPECS Focus 600). The excited photoelectrons were collected by a SPECS 150 hemispherical analyzer at emission and incidence angles of 40° and 60°, respectively. The overall experimental resolution was extracted from Fermi edge analysis of a reference gold sample and resulted in 0.4 eV. The Fermi level position of the reference sample was subtracted to get core level emission in binding energy.

**Survey.** The survey XPS spectrum of the 5-nm–thick Cu on top of graphite after an air exposure of 72 h is shown in Fig. S1. The different peaks are identified and compared to a reference Cu$_2$O spectrum[5]. The survey reveals that there are no other atomic species apart from the expected Cu, O and C.

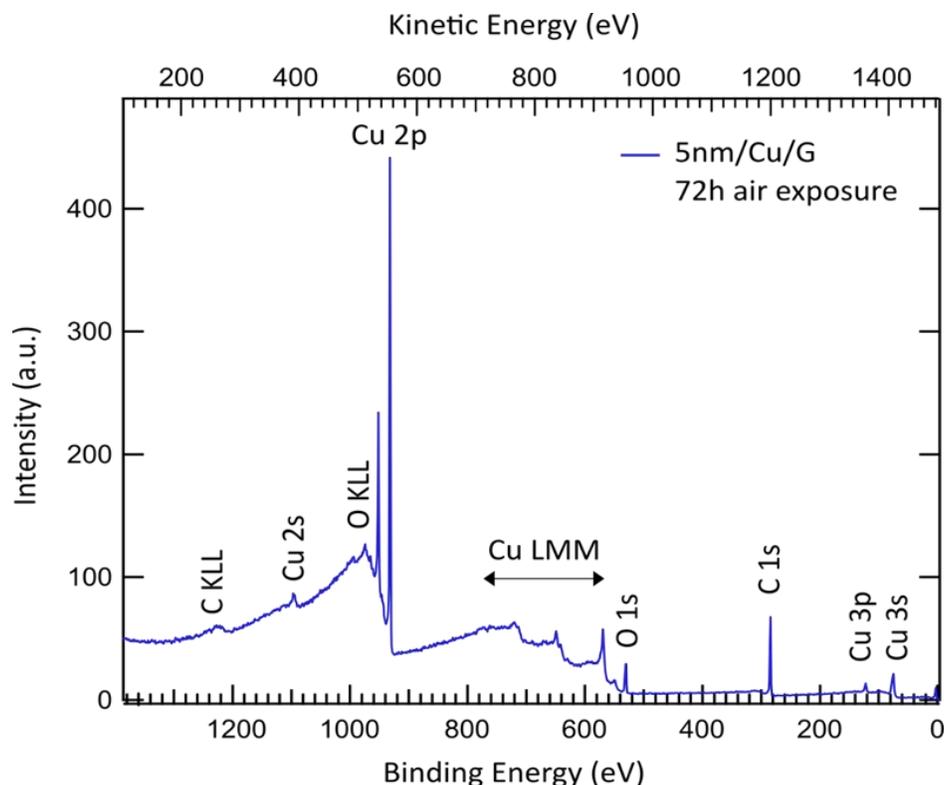

**Fig. S1. Survey.** XPS survey spectrum of a 5-nm-thick Cu film grown on top of graphite and left in air 72 h prior to the measurement. This film was grown exactly under the same experimental conditions as the samples employed to fabricate the devices.

The devices were fabricated using the same Cu coverage and air exposure conditions. For reference, XPS analysis has been performed for Cu deposited on top of exfoliated graphite, while the Cu in the devices was grown on top of exfoliated graphene. According to XPS results, the reference sample of Cu on top of HOPG does not grow as a continuous homogeneous film but follows the inhomogeneities of the HOPG substrate. This leads to the fact that after deposition of a 5-nm-thick Cu, we are still able to detect the C 1s signal of the HOPG substrate. If a 5-nm-thick Cu film would have homogeneously and fully covered the sample, only a very small C signal would have been observed due to the reduced mean free path of the photoelectrons at the approx. 1200 eV kinetic energy. Additionally, the different Cu quantity measured for different sample positions points out that the Cu growth on exfoliated graphite is not homogeneous. For the flatter graphene substrate, however, the Cu growth was more homogenous, as reflected by the experimental AFM measurements (see Fig. 1c of the main text).

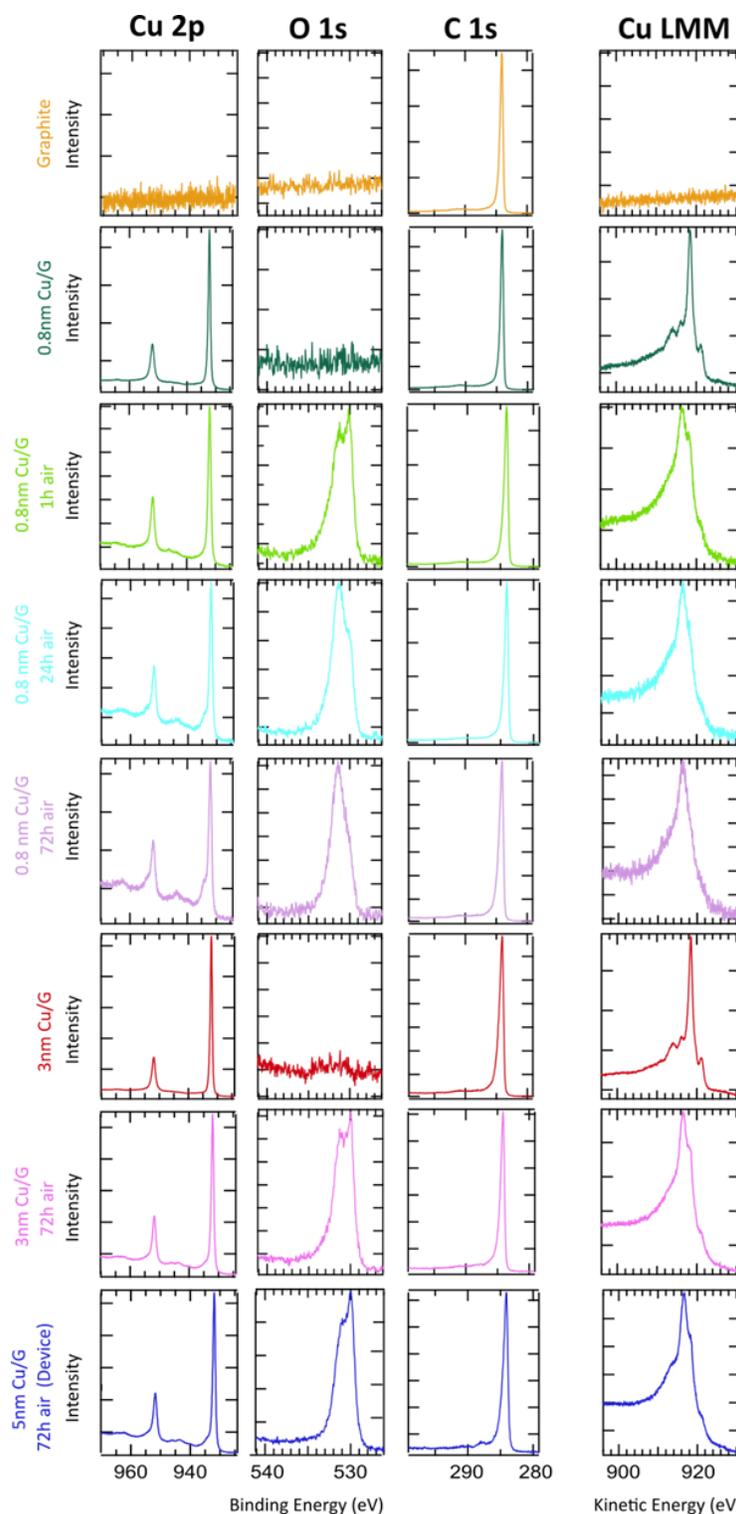

**Fig. S2. XPS summary.** XPS spectra of Cu 2p, Cu LMM, O 1s and C 1s measured for graphite (reference), for a 0.8-nm-thick Cu in UHV, after 1 h, 24 h and 72 h air exposure; for a 3-nm-thick Cu in UHV and after 72 h in air; and for a 5-nm-thick film prepared in the same way as the device after 72 h air exposure.

**Cu 2p levels.** The graphs of Fig. S3 show the Cu 2p core level split into the $2p_{1/2}$ and $2p_{3/2}$ sublevels due to the spin-orbit interaction. In Fig. S3(a), the evolution of the 0.8-nm-thick Cu sample prior and after air exposure is observed. After 1 h air exposure, the film signal decreases significantly due to the presence of air contamination on the surface. The signal continues decreasing with increasing the air exposure time. The binding energies of the Cu $2p_{1/2}$ and $2p_{3/2}$ levels for the 0.8-nm-thick Cu in UHV are 952.0 eV and 932.1 eV, respectively, in agreement with the reported values for metallic Cu(0)[6]. After 1 h air exposure, the 2p levels move slightly to lower binding energies, 951.9 eV and 932.0 eV. There is no further shifts after 24 h and 72 h air exposure times. Although the observed shift to lower binding energies is very small, 0.1-0.2 eV, it reflects the oxidation of the Cu, resulting in $Cu_2O$ (Cu(I) species). For the 0.8-nm-thick sample after 24 h/72 h air exposure, there is a broad peak that emerges around 954.2 eV and 934.3 eV. This emission is related to the appearance of $Cu(OH)_2$. In Fig. S3(b), the sample of 3-nm-thick Cu film grown in UHV and after 72 h air exposure is shown. Similar to the low coverage sample, there is a tiny shift to lower binding energies of the Cu $2p_{1/2}$ and $2p_{3/2}$ peaks after 72 h air exposure and the additional contribution of the $Cu(OH)_2$ is also present for the air exposed sample. Finally, in Fig. S3(c), we compare the spectra for all three samples of different Cu coverages after 72 h air exposure. The binding energies of the core levels are the same. Additionally, we observe that the amount of Cu measured for the 3-nm-thick and the 5-nm-thick is practically the same. This is related to fact that the escape depth of the photoelectrons is smaller than the film thickness and we are reaching the "saturation limit" of XPS.

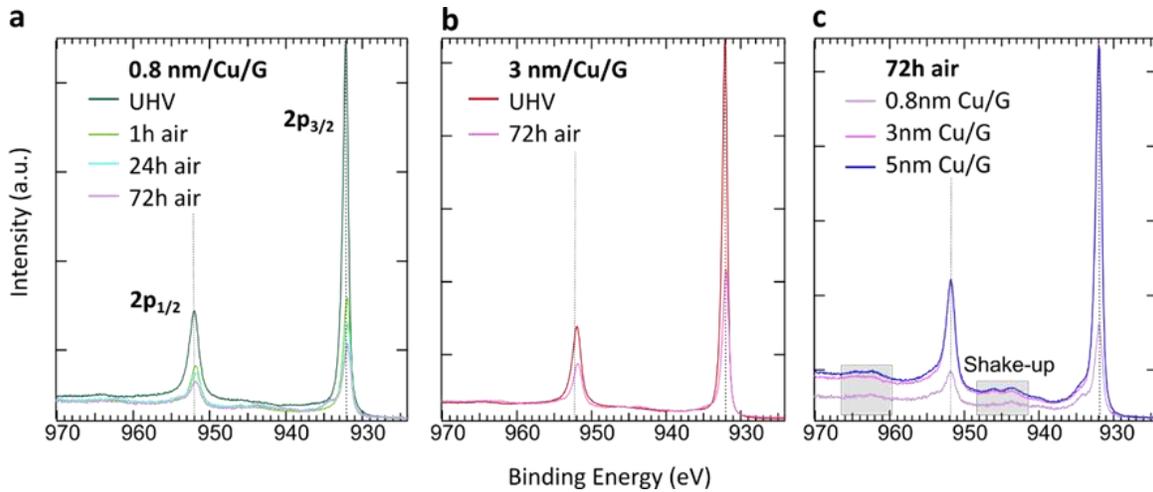

**Fig. S3. Cu 2p.** Cu $2p_{1/2}$ and $2p_{3/2}$ levels for: (a) the low coverage 0.8-nm-thick sample after different air exposure times; (b) the high coverage 3-nm-thick sample in UHV and after 72 h in air, (c) comparison between the different coverages (0.8 nm, 3 nm, and 5 nm of Cu) after 72 h in air.

The Cu quantity on the surface has been estimated as follows. The kinetic energy of the photoemitted electrons of the Cu $2p_{3/2}$ level is $E_{kin} = 553.95$ eV. Therefore, the mean free-path of these electrons[7] is $\lambda = \frac{43}{E^2} + 0.054\sqrt{E} = 1.27$ nm. Comparing the integrals obtained by fitting the peaks with Lorentzian functions and considering that the intensity decreases exponentially as $I = I_0 e^{\frac{-z}{\lambda}}$, we obtained that there is a 2.2 nm difference between the low and high coverages UHV samples, which is in agreement with the QMB reading.

For the air-exposed samples, there are shake-up peaks around 962-964 eV and 944-946 eV, respectively (shadowed areas in Fig. S10(c)). These emissions are related to the $d^9$ configuration of the Cu(II) species, pointing out that there is a small quantity of Cu(II) on the surface. There is no evidence of the presence of CuO in the sample, since 2p peaks related to this compound would be expected at binding energies 933.1 eV and 953 eV. Thus, we can conclude that the Cu(II) species comes from $Cu(OH)_2$. This conclusion is supported by the analysis of the Auger Cu LMM lines and the O 1s spectra.

**Cu LMM.** Figure S4(a) shows the evolution of the Cu $L_3M_{4,5}M_{4,5}$ spectra for the 0.8-nm-thick Cu sample with different air exposure times. For the UHV sample, 6 different peaks can be identified at different kinetic energies, each one with a different intensity, and in complete agreement with the Auger reference spectrum for metallic Cu(0)[8].

| UHV 0.8nm sample | P1 | P2 | P3 | P4 | P5 | P6 |
|---|---|---|---|---|---|---|
| $E_{Kin}$(eV) | 921.2 | 919.6 | 918.5 | 918.0 | 916.1 | 914.1 |

After air exposure, the Auger spectra changes completely. The peaks related to Cu(0) disappear and, in the oxidized spectra, the following new lines appear:

| Air exposed 0.8nm sample | P´1 | P´2 | P´3 |
|---|---|---|---|
| $E_{Kin}$(eV) | 921.5 | 918.1 | 916.4 |

There is not a big difference between the spectra after 1 h, 24 h and 72 h of air exposure, suggesting that the main surface oxidation occurs within the first hour. Additionally, the peaks of the air-exposed sample spectra are quite broad, indicating that there are contributions from Cu(I) and Cu(II) species. Again, a comparison with the Auger lines reported for $Cu_2O$, CuO and $Cu(OH)_2$, indicates the presence of $Cu_2O$ and $Cu(OH)_2$ on the surface.

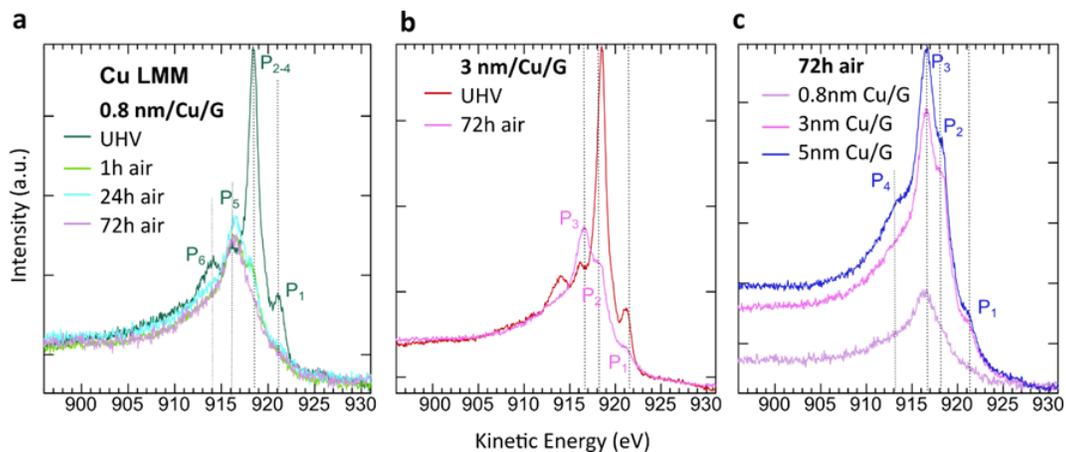

**Fig. S4. Auger Cu LMM.** Cu $L_3M_{4,5}M_{4,5}$ spectra for: (a) the low coverage 0.8-nm-thick sample after different air exposure times; (b) the high coverage 3-nm-thick sample in UHV and after 72 h in air, (c) a comparison between the different coverages (0.8 nm, 3 nm, and 5 nm of Cu) after 72 h in air.

In Fig. S4(b), it is clear that the peaks related to the metallic Cu(0) in the UHV 3-nm-thick Cu sample are completely replaced by at least 3 peaks for the sample left in air for 72 h. A complete oxidation of the Cu species happened. Lastly, a comparison between the three samples with different Cu coverages after 72 h of air exposure is shown in Fig. S4(c). In this case, a fourth peak is more visible for the 5-nm-thick sample with a kinetic energy of 913.5 eV, characteristic of the $Cu_2O$.

**O 1s level.** Figure S5 shows the XPS spectra of the O 1s levels for the different samples. In Fig. S5(a) we observe that there is no O in the exfoliated graphite and in the Cu film grown in UHV conditions. After 1 h of air exposure, two main peaks appear in the spectrum, one at a binding energy of 530.0 eV, which is associated to the $Cu_2O$, and a second one at 531.4 eV, which is related to $Cu(OH)_2$. The hydroxide-related peak is much broader, suggesting that it probably includes other contributions apart from the $Cu(OH)_2$. After 24 h and 72 h of air exposure, the intensity of both peaks increases, becoming the hydroxide peak the most intense one. Table S1 summarizes the employed fit parameters.

Figure S5(e) compares the O 1s spectra of the 0.8-nm-, 3-nm-, and 5-nm-thick Cu samples after 72 h oxidations. For the higher coverages, the most intense contribution arises from $Cu_2O$.

**C 1s level.** Figure S6 shows the evolution of the C 1s peak for the different investigated samples. In Fig. S6(a), we observe how the signal of the C decreases after Cu deposition and, in Fig. S6(b), we see how the signal further decreases with air exposure times. There is no evidence of C oxidation in these spectra. In Fig. S6(c), a comparison between the spectra of the 0.8-nm-, 3-nm-, and 5 nm-thick Cu samples after 72 h of oxidation is shown.

To summarize, from the XPS analysis we observe a complete oxidation of the metallic Cu(0) into Cu(I) and Cu(II) species, arising from $Cu_2O$ and $Cu(OH)_2$.

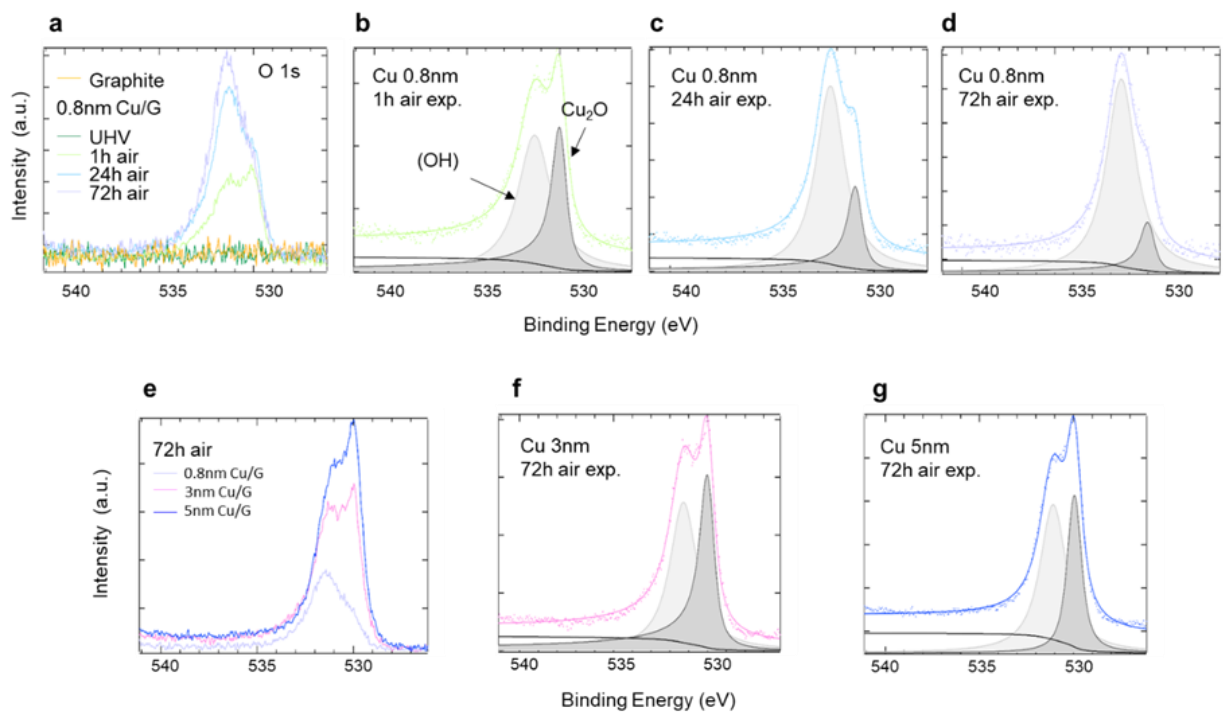

**Fig. S5**. O 1s spectra for: (a) pristine graphite and 0.8-nm-thick Cu in UHV and after different air exposure times; Fits of the O 1s levels for the 0.8-nm-thick sample after (b) 1 h, (c) 24 h and (d) 72 h of air exposure; (e) comparison

between the different coverages after 72 h of air exposure. Fits of the O 1s levels for the (f) 3-nm-thick sample and (g) 5-nm-thick sample after 72 h of air exposure.

| Coverage/Air exposure | Fit Parameters | P1 (Cu$_2$O) | P2 (Hydroxide) |
|---|---|---|---|
| 1 nm after 1 h air exp. | E$_{Bin}$ Position (eV) | 530.0 | 531.4 |
| | FWHM (eV) | 0.29 | 1.04 |
| | Integral | 2.32 | 3.34 |
| 1 nm after 24 h air exp. | E$_{Bin}$ Position (eV) | 530.0 | 531.3 |
| | FWHM (eV) | 0.29 | 0.88 |
| | Integral | 2.27 | 6.71 |
| 1 nm after 72 h air exp. | E$_{Bin}$ Position (eV) | 530.0 | 531.4 |
| | FWHM (eV) | 0.29 | 0.90 |
| | Integral | 1.47 | 7.79 |
| 3 nm after 72 h air exp. | E$_{Bin}$ Position (eV) | 530.0 | 531.3 |
| | FWHM (eV) | 0.29 | 0.82 |
| | Integral | 12.04 | 12.88 |
| 5 nm after 72 h air exp. | E$_{Bin}$ Position (eV) | 530.0 | 531.1 |
| | FWHM (eV) | 0.34 | 0.78 |
| | Integral | 9.28 | 16.62 |

**Table S1.** Main fit parameters of the O1s level.

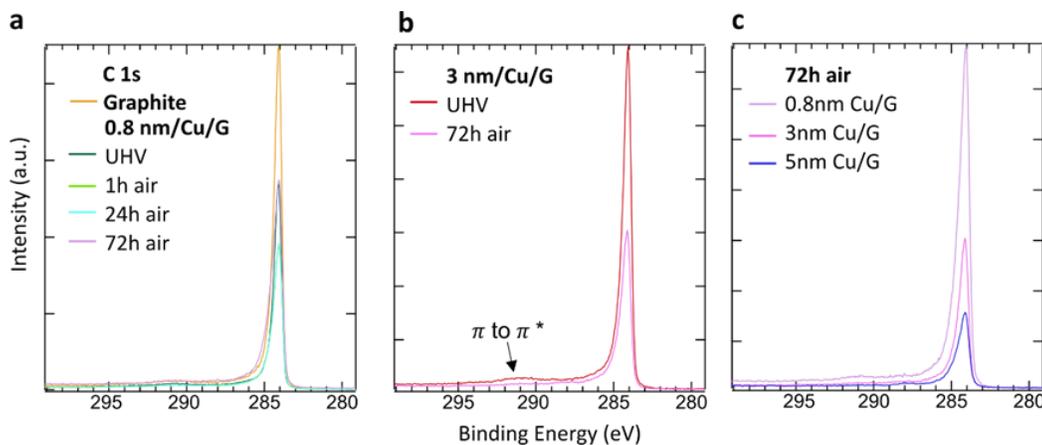

**Fig. S6.** C 1s spectra for: (a) pristine graphite and 0.8-nm-thick Cu in UHV and after different air exposure times; (b) 3-nm-thick Cu in UHV and after 72 h of air exposure; (c) comparison between the different coverages after 72 h of air exposure.

**Supporting Figures.**

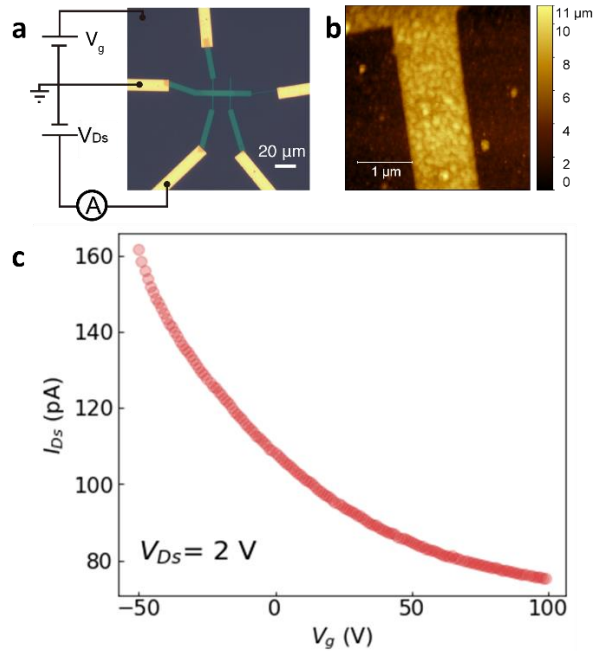

**Fig. S7 Electrical characterization of oxidized copper.** (a) Optical image of a 5-nm-thick Cu Hall bar device grown on a SiO$_2$ substrate, after ambient oxidation and contacted with Ti/Au electrodes. The transport measurement configuration is labeled. (b) AFM image of the oxidized Cu device shows a continuous morphology similar to that grown on graphene and shown in Fig. 1(c) of the main text. (c) Drain-source current ($I_{DS}$) as a function of the back gate voltage ($V_g$) measured at a drain-source voltage ($V_{DS}$) of 2 V and 300 K. One can observe that CuO$_x$ exhibits an insulating behavior. The conductivity of CuO$_x$ is at least seven orders of magnitude smaller than that of the graphene used in our experiment.

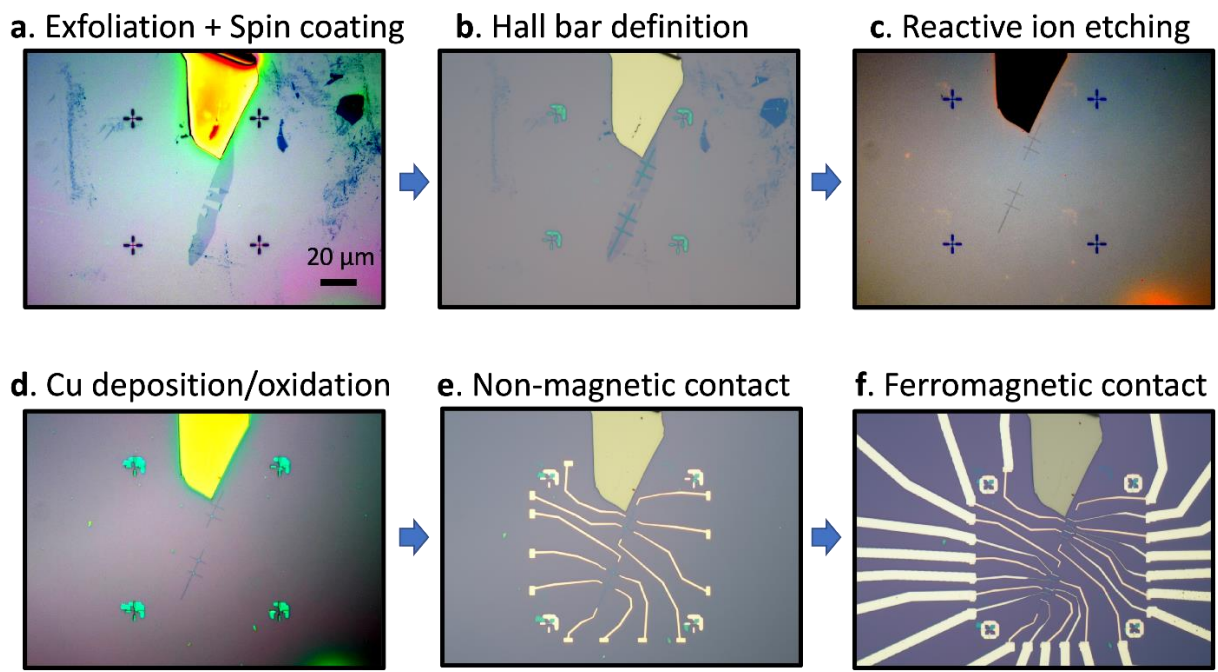

**Fig. S8 Optical image of devices during the fabrication process.** Optical image of Sample 1 and Sample 2 during the fabrication process. (a) Exfoliated single layer graphene on the substrate; (b) Hall bar definition with an Al hard mask; (c) graphene Hall bar after reactive ion etching; (d) Cu deposition on the cross-junction of the Hall bar, followed by ambient oxidation; (e) Non-magnetic Pd/Au contacts deposition; (f) Ferromagnetic $TiO_x$/Co contacts deposition.

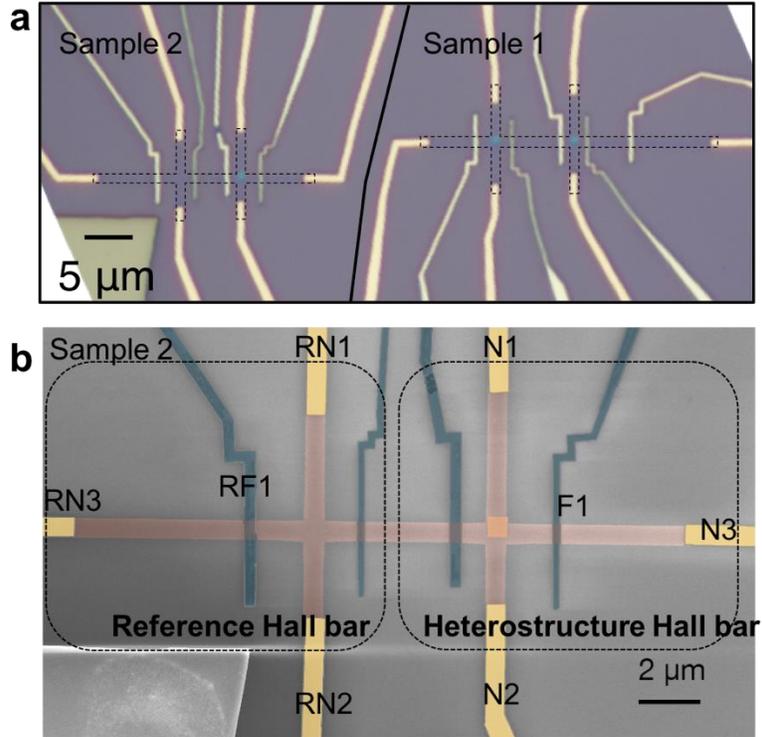

**Fig. S9 Optical and SEM images of Sample 2.** (a) Optical image of Sample 2. The optical image of Sample 1 is also included here, as the two samples are fabricated with the same single layer flake of graphene. The dashed lines indicate the edges of the graphene channel. (b) False-colored SEM image of Sample 2, which includes two non-local Hall bars named reference Hall bar and heterostructure Hall bar, respectively. The reference Hall bar does not have $CuO_x$ covering the cross-junction area. The non-magnetic contact pads (N1 to N3 and RN1 to RN3) and magnetic electrodes (F1 and RF1) are indicated. The data shown in Fig. 4(a) of the main text is measured by applying current from F1 to N3 and detecting the non-local voltage between N1 and N2.

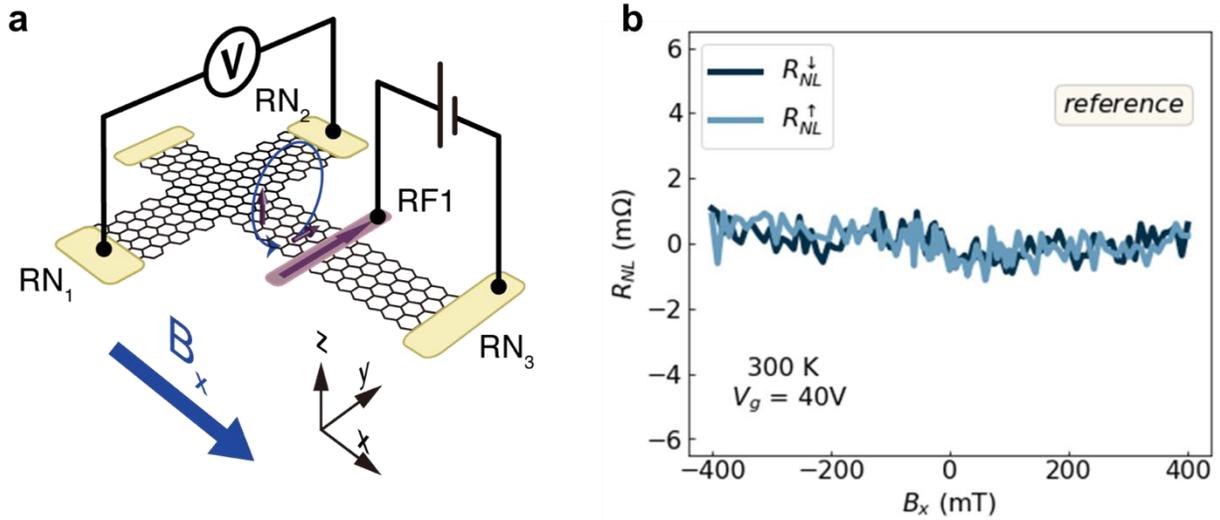

**Fig. S10 Reference non-local Hall bar without CuO$_x$.** (a) Measurement configuration for the reference non-local Hall bar in Sample 2, without CuO$_x$ covering the cross-junction of the Hall bar. (b) Non-local resistance as a function of $B_x$ using configuration in (a), with initial positive ($R_{NL}^\uparrow$, light-blue line) and negative ($R_{NL}^\downarrow$, dark-blue line) magnetization direction of RF1. Data is taken at 300 K and $V_g$ = 40 V, the same condition as the data shown in Fig. 4(a) of the main text. The *y*-axis range is also set to be the same as in Fig. 4(a). We removed the baseline here for a better comparison. No evidence of spin-to-charge conversion is observed when there is no CuO$_x$ on the cross-junction.

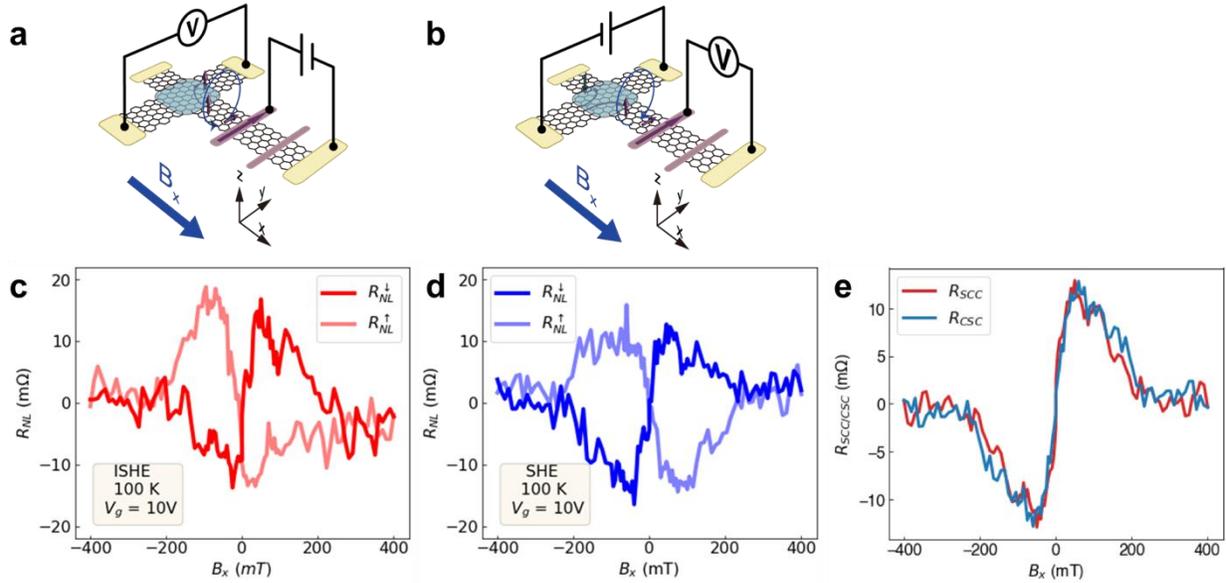

**Fig. S11 Reciprocity between spin Hall effect and inverse spin Hall effect.** Measurement configuration for antisymmetric Hanle precession with (a) spin-to-charge and (b) charge-to-spin conversion at the $CuO_x$-covered graphene and (a) spin injection and (b) detection using a FM electrode, with the results of Sample 1 measured at 100 K and $V_g$ = 10 V shown in (c) and (d), respectively. The baseline is removed and the *y*-axis range is the same for comparison. (e) Net antisymmetric precession signal extracted from the data in (c) (red curve) and (d) (blue curve) by taking $R_{SCC/CSC} = (R_{NL}^{\uparrow} - R_{NL}^{\downarrow})/2$ and antisymmetrizing it. The same amplitude and shape of the precession profile for the SHE and ISHE is obtained, in agreement with the Onsager reciprocity relations for the spin precession process, as has been discussed in the Supplementary Information S5 of Ref. [9].

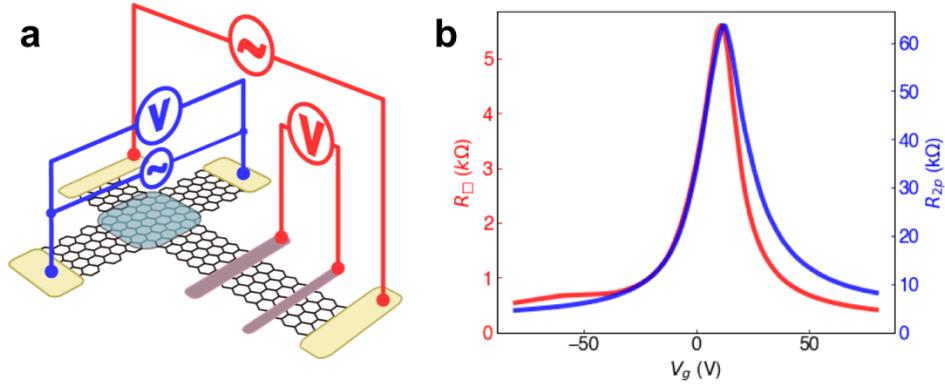

**Fig. S12 Charge transport properties of the pristine graphene and the CuO$_x$-covered graphene.** (a) Measurement configuration for the resistance of the pristine graphene (4-point resistance, in red) and the CuO$_x$-covered graphene at the cross-junction of the Hall bar (2-point resistance, in blue). (b) Square resistance of pristine graphene (red solid line) and 2-point resistance of the CuO$_x$-covered graphene at the cross-junction of the Hall bar (blue solid line) measured in Sample 1. The CNPs for both configurations are the same, indicating that both the CuO$_x$-covered graphene and the pristine graphene exhibit almost the same doping.

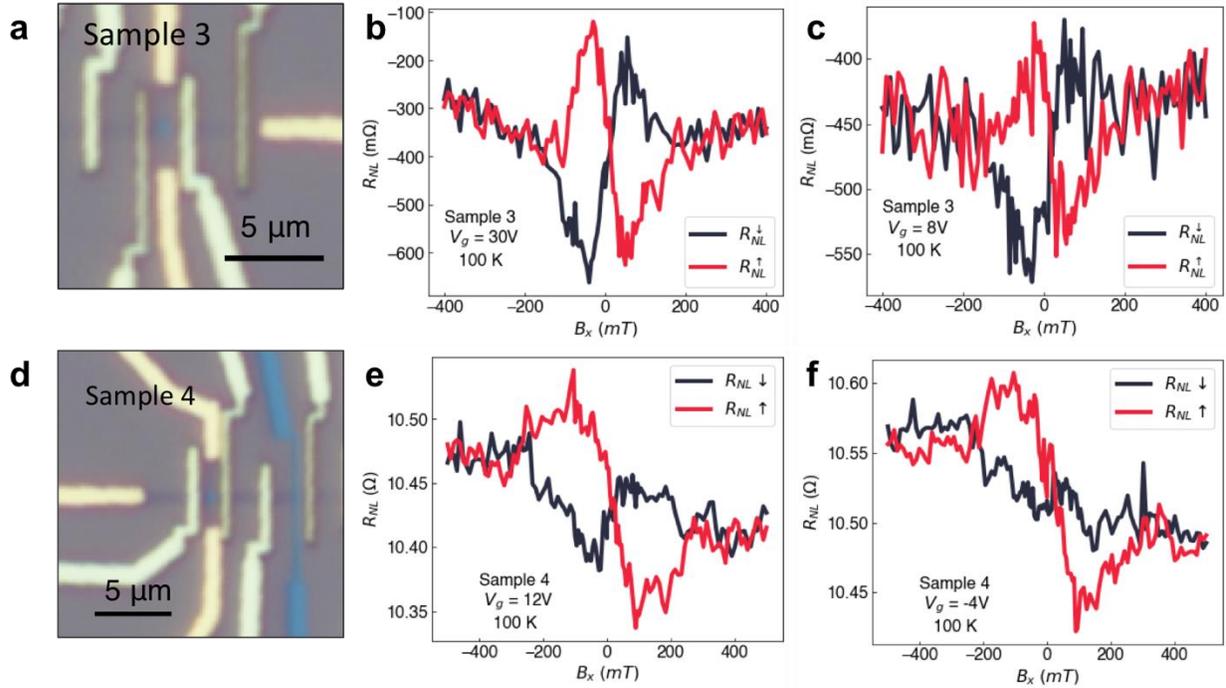

**Fig. S13 Reproducibility.** (a) Optical image of Sample 3. This sample is also fabricated with $CuO_x$ on single layer graphene. (b),(c) Two representative charge-to-spin conversion curves measured at 100 K with $V_g$ = 30 V and 8 V, respectively. The signal shows a clear antisymmetric component, indicating the detection of the SHE in the $CuO_x$/graphene heterostructure. (d) Optical image of Sample 4. The sample is fabricated with $CuO_x$ on trilayer graphene. (e),(f) Two representative charge-to-spin conversion curves measured at 100 K with $V_g$ = 12 V and −4 V, respectively.

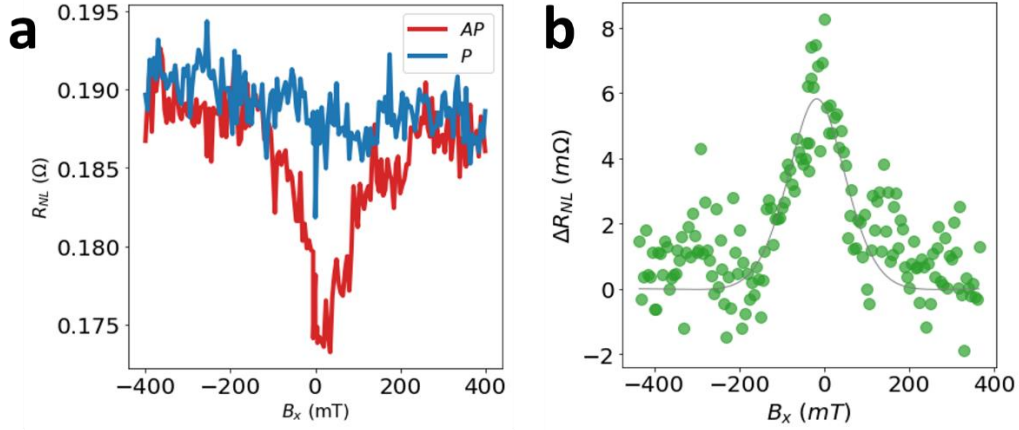

**Fig. S14 Cross-section Hanle precession.** (a) Non-local resistance across the CuO$_x$/graphene region as a function of $B_x$ measured at 300 K and $V_g$ = 0 V, with the FM electrodes set in a parallel ($R_{NL}^P$, blue line) and antiparallel ($R_{NL}^P$, red line) configuration. (b) Net symmetric Hanle precession signal extracted from the two curves in (a) by taking $\Delta R_{NL} = (R_{NL}^P - R_{NL}^{AP})/2$. Here, no evidence of spin lifetime anisotropy is observed. In a TMDC/graphene system, the spin lifetime anisotropy stems from the valley-Zeeman interaction resulting from the TMD layer. However, unlike the existence of valleys in TMDC crystals, the Cu oxide is deposited by thermal evaporation in an amorphous state. In this regard, no spin lifetime anisotropy was expected in this system.

# SUPPORTING REFERENCES